\documentclass[11pt,letter,onecolumn]{emulateapj}
\usepackage{amsmath}
\usepackage{graphicx}
\usepackage{epstopdf}

\newcommand{\QU}{$Q$-$U$\,}
\DeclareMathOperator{\sign}{sign}

\shorttitle{Advanced Polarization Diagnostics}
\shortauthors{Herron et al.}
\begin{document}

\title{Advanced Diagnostics for the Study of Linearly Polarized Emission. I: Derivation}

\author{C.~A.~Herron\altaffilmark{1}, B.~M.~Gaensler\altaffilmark{2,1}, G.~F.~Lewis\altaffilmark{1}, N.~M.~McClure-Griffiths\altaffilmark{3}}

\altaffiltext{1}{Sydney Institute for Astronomy, School of Physics, A28, The University
of Sydney, NSW, 2006, Australia; C.Herron@physics.usyd.edu.au}

\altaffiltext{2}{Dunlap Institute for Astronomy and Astrophysics, University of Toronto, 50 St. George Street,
Toronto, Ontario, M5S 3H4, Canada}

\altaffiltext{3}{Research School of Astronomy and Astrophysics, The Australian National University, Canberra, ACT 2611, Australia}

\date{\today}


\begin{abstract}
Linearly polarized emission is described, in general, in terms of the Stokes parameters $Q$ and $U$, from which the polarization intensity and polarization angle can be determined. Although the polarization intensity and polarization angle provide an intuitive description of the polarization, they are affected by the limitations of interferometric data, such as missing single-dish data in the u-v plane, from which radio frequency interferometric data is visualized. To negate the effects of these artefacts, it is desirable for polarization diagnostics to be rotationally and translationally invariant in the $Q$-$U$ plane. One rotationally and translationally invariant quantity, the polarization gradient, has been shown to provide a unique view of spatial variations in the turbulent interstellar medium when applied to diffuse radio frequency synchrotron emission. In this paper we develop a formalism to derive additional rotationally and translationally invariant quantities. We present new diagnostics that can be applied to diffuse or point-like polarized emission in any waveband, including a generalization of the polarization gradient, the polarization directional curvature, polarization wavelength derivative, and polarization wavelength curvature. In Paper II we will apply these diagnostics to observed and simulated images of diffuse radio frequency synchrotron emission.

\end{abstract}
\section{Introduction} 
\label{intro} 
Observations of polarized emission are capable of providing unique insights on many astrophysical processes, which cannot be obtained from an analysis of the total intensity emission alone. For example, the Zeeman splitting of atomic and molecular spectral lines provides a method to measure the magnetic field in cold neutral or molecular gas \citep{Crutcher2010, Robishaw2015}. Polarized thermal dust emission has been used to improve models of the large-scale Galactic magnetic field \citep{Planck2016b}, and can provide insights on turbulent driving mechanisms in the interstellar medium (ISM, \citealt{Caldwell2016}).

At radio wavelengths, observations of polarized synchrotron emission from point sources and its associated Faraday rotation have been used to study the large-scale Galactic \citep{Oppermann2015} and pan-Magellanic \citep{Kaczmarek2017} magnetic fields, and observations of polarized diffuse emission have been used to study the local ISM \citep{Lenc2016, VanEck2017} and large Galactic structures \citep{Sun2015, Hill2017}. Radio observations can also be used to study the magnetic fields of other galaxies. For example, \cite{OSullivan2013} probed the thermal gas, magnetic field and radio-emitting electrons in the lobes of Centaurus A, coherent magnetic fields in galaxies \citep{Fletcher2011} and merging galaxies \citep{Basu2017} have been detected, and fluctuations of the observed polarized emission from M51 suggest that superbubbles and Parker instabilities drive the interstellar turbulence in that galaxy \citep{Mao2015}. 

In the future, polarimetry is set to play an even more significant role in improving our understanding of astrophysical processes, as the Square Kilometre Array will provide detailed polarimetric information on millions of point sources, and large areas of diffuse emission (see \citealt{Gaensler2015} and \citealt{Johnston2015} for an overview).

As the measurement of polarized emission has significant potential to provide unique insights, it is important to ensure that all of the information encoded by the polarization state of the emission is extracted, in a manner that is robust to observational artefacts and amenable to statistical analysis. For emission that is linearly polarized, the intuitive quantities to measure are the polarization intensity $P$, and the polarization angle $\psi$, measured anti-clockwise from North (see \citealt{Hamaker1996} for an interpretation of the conventions established by the \citealt{IAU1974}).

Rather than use $P$ and $\psi$, we can also use the Stokes parameters $Q$ and $U$ to provide a complete description of linearly polarized emission. The Stokes parameters are orthogonal decompositions of the linear polarization, such that $Q$ measures vertical and horizontal polarization, and $U$ measures the polarization along diagonals at $45^{\circ}$ to the vertical and horizontal (see \citealt{Stokes1852}, \citealt{Gardner1966}, and \citealt{Saikia1988} for introductions to the Stokes parameters). From $Q$ and $U$, the polarization intensity and polarization angle are calculated according to
\begin{equation}
P = \sqrt{Q^2 + U^2}, \text{ and } \psi = \frac{1}{2} \arctan \frac{U}{Q}.
\end{equation}
We can then define the complex polarization $\boldsymbol{P} = Q + iU$, such that the polarization intensity $P = |\boldsymbol{P}|$, and the azimuthal angle in the complex \QU plane is $2\psi$.

Although the polarization intensity and polarization angle are intuitive quantities, they suffer from the limitations of interferometric data, and poor statistical properties, that complicate the link between these quantities and astrophysical processes. For instance, while $Q$ and $U$ exhibit Gaussian noise properties, this is not true for $P$ and $\psi$. The noise in $Q$ and $U$ is squared when calculating $P$, having a Ricean distribution, and this causes the observed polarization intensity to be biased towards larger values \citep{Wardle1974, Simmons1985}. This has been a key topic recently, with \cite{Montier2015a, Montier2015b}, \cite{Vidal2016} and \cite{Muller2017} examining how noise and noise correlations affect $P$ and $\psi$, and developing methods to remove the bias from polarization intensity.

A limitation of interferometric data that can influence the measured values of $P$ and $\psi$ is missing data in the u-v plane, from which radio frequency interferometric data is visualized. If short-baseline data, or single-dish data, are missing from the u-v plane, then the produced image will not display emission from large-scale structures. We can consider the effect of missing short-baseline data as causing a translation in the \QU plane, because we would need to add the Stokes $Q$ and $U$ of the large-scale structure to the observed $Q$ and $U$, in order to have a complete image. Hence, we desire polarimetric diagnostics to be translationally invariant in the \QU plane to avoid the effect of missing short-baseline data. Rotation in the \QU plane could be caused by rotating the coordinate system used to define the polarization angle, as this would change the measured values of $\psi$. Any physically meaningful quantity should not depend on the coordinate system used to measure it, and so we also desire polarimetric quantities that are rotationally invariant in the \QU plane.

A quantity that is rotationally and translationally invariant in the \QU plane is the `polarization gradient', $|\nabla {\boldsymbol{P}}|$, derived by \cite{Gaensler2011}. The amplitude of the polarization gradient is given by
\begin{equation}
{|\nabla {\boldsymbol{P}}|}={\sqrt{\biggl(\frac{\partial Q}{\partial x}\biggr)^2
+\biggl(\frac{\partial U}{\partial x}\biggr)^2
+\biggl(\frac{\partial Q}{\partial y}\biggr)^2
+\biggl(\frac{\partial U}{\partial y}\biggr)^2},}
\label{p_grad}
\end{equation}
where $x$ and $y$ are the Cartesian axes of the image plane. Eq. \ref{p_grad} illustrates that the polarization gradient traces spatial changes in Stokes $Q$ and $U$, and hence spatial changes of the complex polarization as a whole. The angle that the polarization gradient makes with the $x$ axis of the image is given by
\begin{equation}
{\arg (\nabla {\boldsymbol{P}})}=\arctan \Bigg[{ \sign \biggl(\frac{\partial Q}{\partial x} \frac{\partial Q}{\partial y} + \frac{\partial U}{\partial x} \frac{\partial U}{\partial y} \biggr) \sqrt{\biggl(\frac{\partial Q}{\partial y}\biggr)^2
+\biggl(\frac{\partial U}{\partial y}\biggr)^2} \Bigg/ \sqrt{\biggl(\frac{\partial Q}{\partial x}\biggr)^2
+\biggl(\frac{\partial U}{\partial x}\biggr)^2} \Bigg].}
\label{p_grad_arg}
\end{equation}

\cite{Gaensler2011} calculated the amplitude of the polarization gradient for the Southern Galactic Plane Survey \citep{McClure2001}, and found that it traced spatial variations in the warm-ionized medium caused by vorticity, shear, and shocks, providing a unique view of interstellar turbulence. \cite{Burkhart2012} calculated the amplitude of the polarization gradient for mock observations of synchrotron emission propagating through a turbulent magnetoionic medium, and found that statistics of polarization gradient structures, such as the genus, were sensitive to the regime of turbulence. The polarization gradient has been applied to observations by \cite{Iacobelli2014}, \cite{Sun2014} and \cite{Herron2017}, to constrain the sonic Mach number of observed magnetoionic turbulence, and \cite{Robitaille2015} found that there were different networks of filaments in the images of the amplitude of the polarization gradient on different angular scales in a field of the Canadian Galactic Plane Survey \citep{Landecker2010}. \cite{Robitaille2017} also compared the polarization gradient to the E- and B-modes \citep{Zaldarriaga1997} of diffuse synchrotron emission in the S-band Polarization All Sky Survey \citep{Carretti2010, Carretti2013}, and found that the two had similar properties, and provide complementary information on observed magnetoionic turbulence.

Although new rotationally and translationally invariant quantities have yet to be examined, there has been progress in developing promising diagnostics of turbulence. \cite{Soler2013} developed the Histogram of Relative Orientations, and used it to study the alignment of the magnetic field with filaments of molecular gas. \cite{Lazarian2016} introduced two new methods that use fluctuations in synchrotron polarization to study magnetoionic turbulence in the ISM. The first method, Polarization Spatial Analysis, provides the spectrum of magnetic fluctuations, and is sensitive to the ratio of the regular to the random magnetic field \citep{Lee2016}. The second method, Polarization Frequency Analysis, provides information on the statistics of the magnetic field and Faraday rotation \citep{Zhang2016}.

In this paper we derive new quantities that are rotationally and translationally invariant in the \QU plane. In Section \ref{framework} we introduce the framework that we use to derive new invariant quantities. In Sections \ref{1Space} and \ref{2Space} we derive invariant quantities that involve the first and second order spatial derivatives of Stokes $Q$ and $U$ respectively. We derive invariant quantities that involve the first and second order derivatives of $Q$ and $U$ with respect to wavelength in Sections \ref{1Wave} and \ref{2Wave} respectively, and in Appendix \ref{Mixed} we derive quantities that depend on spatial and spectral derivatives of $Q$ and $U$. We discuss potential applications of the new invariant quantities in Section \ref{discuss}, and conclude in Section \ref{concl}. Although we focus on diffuse, radio-frequency synchrotron emission throughout this paper, we emphasize that the diagnostics derived in this paper can be applied to any linearly polarized emission, in any waveband, and the interpretation of the diagnostics will depend upon the object observed. In Paper II, we calculate these diagnostics for observed and simulated radio synchrotron emission arising from the turbulent, magnetoionic ISM.

\section{Theoretical Framework}
\label{framework}
\begin{figure*}
\begin{center}
\includegraphics[scale=0.5]{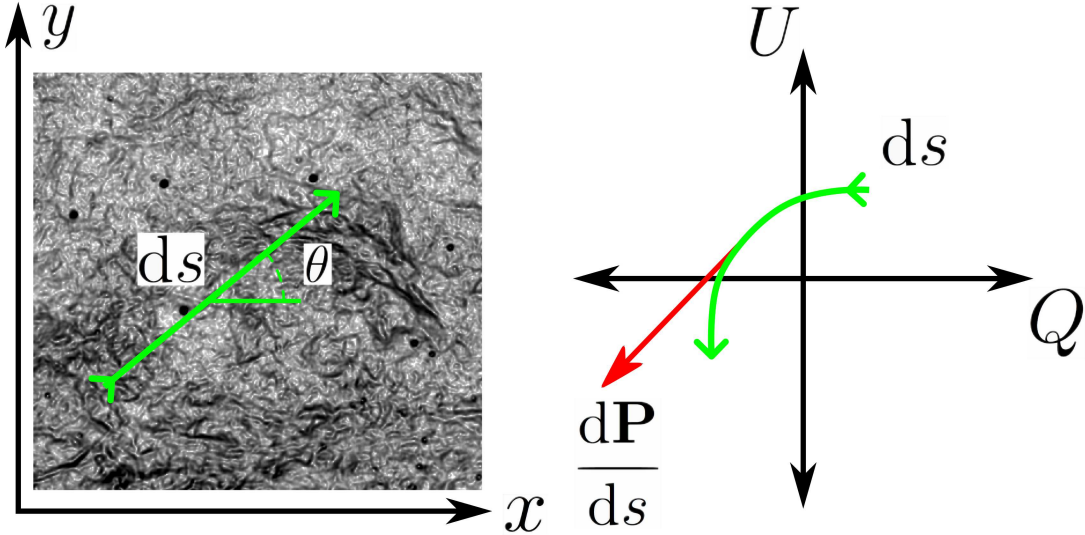}
\caption{A diagram illustrating the calculation of the polarization directional derivative. The image plane is on the left, and we calculate the derivative of the polarization along a path, in the direction determined by the angle $\theta$. The path in the image plane corresponds to a path in the \QU plane, and we calculate the speed at which we traverse this path. The background image is of polarization gradients from a portion of the Canadian Galactic Plane Survey, see \cite{Herron2017} for more information.}
\label{direc_deriv_path}
\end{center}
\end{figure*}

To derive quantities that are rotationally and translationally invariant in the \QU plane, we consider a region of the sky that has been imaged over a range of wavelengths. We define the Cartesian $x$ and $y$ axes in this image, which lies in the plane of the sky, as shown in Fig. \ref{direc_deriv_path}. At a given wavelength $\lambda$, Stokes $Q$ and $U$ will vary across this image, so we consider $Q$ and $U$ as functions of space and wavelength, i.e. $Q = Q(x,y,\lambda^2)$ and $U = U(x,y,\lambda^2)$. Note that we consider $Q$ and $U$ as functions of $\lambda^2$, rather than of $\lambda$, because a common cause of wavelength dependent polarization is the Faraday rotation of polarized emission as it propagates through a magnetoionic medium, and this effect is proportional to $\lambda^2$. For the equations presented in this paper, it is also valid to replace $\lambda^2$ with $\lambda$, or with the frequency $\nu$ of the emission.

We consider moving along a line element in the image plane, of length $\textrm{d}s$, oriented at angle $\theta$ relative to the $x$-axis, as shown in Fig. \ref{direc_deriv_path}. As we move along this path, we trace out a corresponding path in the \QU plane, parametrized by the distance in the image plane, $s$. We can describe this path by a vector in the complex plane, $\boldsymbol{P}(s) = (Q(s), U(s))$. It is intuitively clear that the speed at which we traverse the path in the \QU plane, and the shape of this path, are invariant under rotations and translations of the \QU plane.

\begin{figure*}
\begin{center}
\includegraphics[scale=0.5]{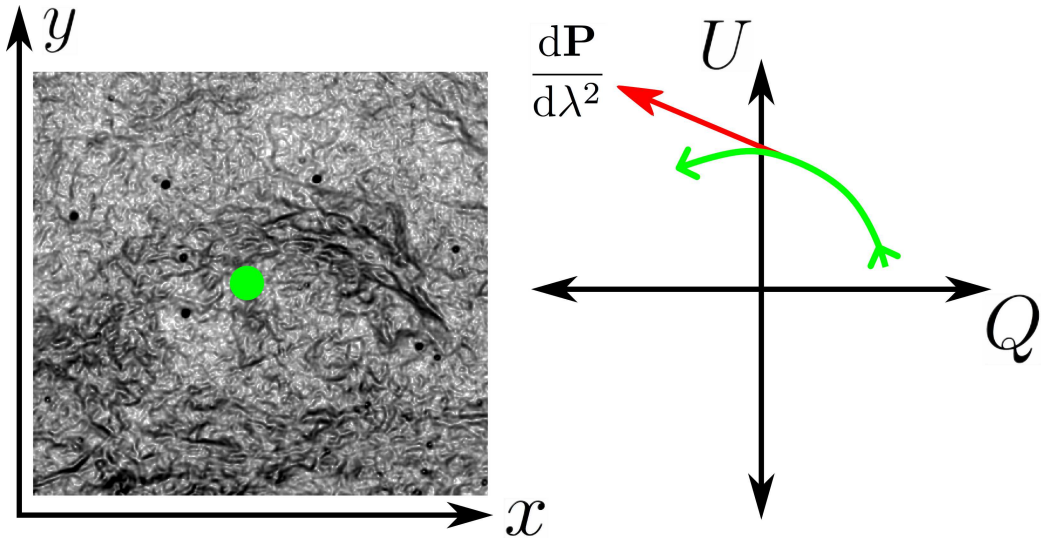}
\caption{A diagram illustrating the calculation of the polarization wavelength derivative. The image plane is on the left, and we consider the polarization at a pixel in the image, marked by the green dot. As the wavelength changes, a path is traversed in the \QU plane, and we calculate the speed at which we traverse this path. The background image is of polarization gradients from a portion of the Canadian Galactic Plane Survey, see \cite{Herron2017} for more information.}
\label{wav_deriv_path}
\end{center}
\end{figure*}

The rate of change of the polarization vector with distance in the image plane is
\begin{equation}
\frac{\partial \boldsymbol{P}}{\partial s} = \biggl( \frac{\partial Q}{\partial s}, \frac{\partial U}{\partial s}\biggr), \label{pol_velocity_space}
\end{equation}
where the derivatives of $Q$ and $U$ with respect to $s$ can be found via the chain rule:
\begin{equation}
\frac{\partial Q}{\partial s} =  \frac{\partial Q}{\partial x} \frac{\partial x}{\partial s} +  \frac{\partial Q}{\partial y} \frac{\partial y}{\partial s}, \text{ and }
\frac{\partial U}{\partial s} = \frac{\partial U}{\partial x} \frac{\partial x}{\partial s} +  \frac{\partial U}{\partial y} \frac{\partial y}{\partial s}.
\end{equation}
By considering infinitesimal lengths in the image plane, we also have
\begin{equation}
\frac{\partial x}{\partial s} = \cos \theta \text{, and } \frac{\partial y}{\partial s} = \sin \theta.
\end{equation}
Having established the framework for spatial derivatives of polarization, we now consider the analogous case of derivatives with respect to wavelength. For this case, we consider a single point in the image plane, shown by a green dot in Fig. \ref{wav_deriv_path}. As we change the observing wavelength, a curve is traced in the Q-U plane, and the speed and curvature of this path should be invariant quantities.

The rate of change of the polarization with respect to wavelength squared is then given by 
\begin{equation}
\frac{\partial \boldsymbol{P}}{\partial \lambda^2} = \biggl( \frac{\partial Q}{\partial \lambda^2}, \frac{\partial U}{\partial \lambda^2}\biggr). \label{pol_velocity_wav}
\end{equation}
To test whether a diagnostic is rotationally and translationally invariant in the \QU plane, we consider rotating the \QU plane anti-clockwise by angle $\phi$, and then translating the \QU plane by $a+ib$, so that the new complex polarization\footnote{Equivalently, it is possible to translate first, and then rotate, to test that a diagnostic is rotationally and translationally invariant in a similar fashion.} is given by $\boldsymbol{P}^*(x,y,\lambda^2) = Qe^{i\phi} + iUe^{i\phi} +a+ib$. By collecting real and imaginary parts, the values of Stokes $Q$ and $U$ in the transformed plane, $Q^*$ and $U^*$, are given by
\begin{equation}
Q^* = Q \cos \phi - U \sin \phi + a, \text{ and }
U^* = Q \sin \phi + U \cos \phi + b. \label{rot_pol}
\end{equation}
We perform calculations on the transformed Stokes parameters, by substituting them for $Q$ and $U$ in Eq. \ref{pol_velocity_space} and Eq. \ref{pol_velocity_wav}. A diagnostic is then rotationally and translationally invariant in the \QU plane if it is independent of $\phi$, $a$, and $b$.

We note that it is possible to generalize this framework to include circular polarization, by considering the polarization vector as $\boldsymbol{P} = (Q,U,V)$, and three-dimensional rotations and translations, however this is beyond the scope of this paper.

\section{First Order Spatial Derivatives}
\label{1Space}
We test our framework for calculating polarization diagnostics by attempting to derive the polarization gradient. We start with Eq. \ref{pol_velocity_space} for spatial changes of complex polarization, and calculate the amplitude of this vector, to find the total rate of change of polarization in a specified direction. This amplitude is
\begin{equation}
\bigg|\frac{\partial \boldsymbol{P}}{\partial s} \bigg| = \sqrt{\cos^2 \theta \biggl( \biggl( \frac{\partial Q}{\partial x} \biggr)^2 + \biggl( \frac{\partial U}{\partial x} \biggr)^2 \biggr) + 2 \cos \theta \sin \theta \biggl( \frac{\partial Q}{\partial x}  \frac{\partial Q}{\partial y} + \frac{\partial U}{\partial x} \frac{\partial U}{\partial y} \biggr) + \sin^2 \theta \biggl(  \biggl( \frac{\partial Q}{\partial y} \biggr)^2 +  \biggl( \frac{\partial U}{\partial y} \biggr)^2 \biggr) }. \label{direc_deriv}
\end{equation}
In Eq. \ref{direc_deriv}, $\theta$ is a diagnostic parameter that can be specified to calculate the rate of change of the complex polarization vector at a position in an image, in a specified direction. We call Eq. \ref{direc_deriv} the `polarization directional derivative', and it is rotationally and translationally invariant in the \QU plane. 

We interpret the polarization gradient as defined in Eqs. \ref{p_grad} and \ref{p_grad_arg} as the maximum rate of change of the complex polarization vector, and so we attempt to derive the polarization gradient by maximizing the polarization directional derivative over $\theta$. We first consider the simplest case, where the observed polarization has uniform polarization intensity across the image. This case may occur if there is a sheet of polarized emission propagating through a Faraday rotating medium, for instance, and we refer to it as the `backlit' case. In this case, we substitute $Q = P \cos 2 \psi$ and $U = P \sin 2 \psi$ into Eq. \ref{direc_deriv}, and find that the directional derivative is given by
\begin{equation}
\bigg|\frac{\partial \boldsymbol{P}}{\partial s} \bigg|_{\text{backlit}} = 2P \bigg| \frac{\partial \psi}{\partial x} \cos \theta + \frac{\partial \psi}{\partial y} \sin \theta \bigg|, \label{backlit_direc_deriv}
\end{equation}
and it is maximized for an angle $\theta_{\text{max,backlit}}$ given by
\begin{equation}
\tan \theta_{\text{max,backlit}} = \frac{\partial \psi}{\partial y} \bigg/ \frac{\partial \psi}{\partial x} = \frac{\partial Q}{\partial y} \bigg/ \frac{\partial Q}{\partial x} = \frac{\partial U}{\partial y} \bigg/ \frac{\partial U}{\partial x}. \label{backlit_ang_max}
\end{equation}
The maximum amplitude of the directional derivative for the case of backlit emission is then given by
\begin{equation}
\bigg|\frac{\partial \boldsymbol{P}}{\partial s} \bigg|_{\text{max,backlit}} = \sqrt{4P^2 \biggl( \frac{\partial \psi}{\partial x} \biggr)^2 + 4P^2 \biggl( \frac{\partial \psi}{\partial y} \biggr)^2} = \sqrt{\biggl(\frac{\partial Q}{\partial x}\biggr)^2 +\biggl(\frac{\partial U}{\partial x}\biggr)^2 +\biggl(\frac{\partial Q}{\partial y}\biggr)^2 +\biggl(\frac{\partial U}{\partial y}\biggr)^2} = |\nabla {\boldsymbol{P}}|, \label{backlit_max_direc_deriv}
\end{equation}
and so for the case of backlit emission, the maximum amplitude of the directional derivative is the same as the amplitude of the polarization gradient, as expected. However, our formula for the angle that maximizes the directional derivative differs from the formula given for the angle that the polarization gradient makes with the $x$ axis of the image, given by Eq. \ref{p_grad_arg}. We will discuss possible reasons for this discrepancy at the end of this subsection.

For the most general case where both the polarization intensity and polarization angle vary across an image, we find that the polarization directional derivative is maximized for an angle $\theta_{\text{max}}$ that satisfies both
\begin{align}
\cos 2\theta_{\text{max}} &= \frac{ - \bigl( \bigl( \frac{\partial Q}{\partial y} \bigr)^2 - \bigl( \frac{\partial Q}{\partial x} \bigr)^2 + \bigl( \frac{\partial U}{\partial y} \bigr)^2 - \bigl( \frac{\partial U}{\partial x} \bigr)^2 \bigr) }{ \sqrt{ \bigl( \bigl( \frac{\partial Q}{\partial x} \bigr)^2 + \bigl( \frac{\partial Q}{\partial y} \bigr)^2 + \bigl( \frac{\partial U}{\partial x} \bigr)^2 + \bigl( \frac{\partial U}{\partial y} \bigr)^2 \bigr)^2 - 4 \bigl( \frac{\partial Q}{\partial x} \frac{\partial U}{\partial y} - \frac{\partial Q}{\partial y} \frac{\partial U}{\partial x} \bigr)^2 } } \text{, and} \label{direc_theta_max_cos}\\
\sin 2\theta_{\text{max}} &= \frac{ 2 \bigl( \frac{\partial Q}{\partial x} \frac{\partial Q}{\partial y} + \frac{\partial U}{\partial x}  \frac{\partial U}{\partial y} \bigr) }{ \sqrt{ \bigl( \bigl( \frac{\partial Q}{\partial x} \bigr)^2 + \bigl( \frac{\partial Q}{\partial y} \bigr)^2 + \bigl( \frac{\partial U}{\partial x} \bigr)^2 + \bigl( \frac{\partial U}{\partial y} \bigr)^2 \bigr)^2 - 4 \bigl( \frac{\partial Q}{\partial x} \frac{\partial U}{\partial y} - \frac{\partial Q}{\partial y} \frac{\partial U}{\partial x} \bigr)^2 } }  \text{,} \label{direc_theta_max_sin}
\end{align}
and the maximum amplitude of the directional derivative is given by
\begin{align}
\bigg| \frac{\partial \boldsymbol{P}}{\partial s} \bigg|_{\text{max}} &= \Biggl[ \frac{1}{2} \biggl( \biggl(\frac{\partial Q}{\partial x} \biggr)^2 + \biggl(\frac{\partial U}{\partial x} \biggr)^2 + \biggl(\frac{\partial Q}{\partial y} \biggr)^2 + \biggl(\frac{\partial U}{\partial y} \biggr)^2  \biggr) + \nonumber \\ 
 & \frac{1}{2} \sqrt{\biggl( \biggl(\frac{\partial Q}{\partial x} \biggr)^2 + \biggl(\frac{\partial U}{\partial x} \biggr)^2 + \biggl(\frac{\partial Q}{\partial y} \biggr)^2 + \biggl(\frac{\partial U}{\partial y} \biggr)^2 \biggr)^2 - 4 \biggl( \frac{\partial Q}{\partial x} \frac{\partial U}{\partial y} - \frac{\partial Q}{\partial y} \frac{\partial U}{\partial x} \biggr)^2  } \Biggr]^{1/2}.  \label{direc_max}
\end{align}
Both the maximum amplitude of the directional derivative and $\theta_{\text{max}}$ are rotationally and translationally invariant in the \QU plane. From Eq. \ref{direc_max}, it is clear that the maximum amplitude of the directional derivative differs from the amplitude of the polarization gradient as given in Eq. \ref{p_grad}. The difference lies in the final term beneath the square root, which can be interpreted as the determinant of the Jacobian that defines the transformation between the image and \QU planes, or alternatively, as the amplitude of the cross product of the gradient of $Q$ and the gradient of $U$. If this term is zero, then the maximum amplitude of the directional derivative is the same as the amplitude of the polarization gradient. This implies that the amplitude of the polarization gradient is only equal to the maximum change of polarization for the special case of backlit emission, and so we will refer to the maximum amplitude of the directional derivative as the `generalized polarization gradient'. We will discuss the interpretation of the polarization gradient further in Section \ref{rad_tang_space}. As the generalized polarization gradient is very similar to the polarization gradient, we believe that the generalized polarization gradient will similarly trace changes in the complex polarization that may be caused by vorticity, shear, or shocks, when applied to diffuse synchrotron emission, and we test this proposal in Paper II. From Eqs. \ref{direc_theta_max_cos} and \ref{direc_theta_max_sin}, it is clear that $\theta_{\text{max}}$ differs from $\arg (\nabla {\boldsymbol{P}})$. We believe that the differences between the generalized polarization gradient and the polarization gradient derived by \cite{Gaensler2011} arise because of how the transformation between the image and \QU planes is treated in our framework, which was not considered by \cite{Gaensler2011}. This caused \cite{Gaensler2011} to implicitly assume that the gradient of $Q$ and the gradient of $U$ were in the same direction in the image plane.

In addition to calculating the maximum amplitude of the directional derivative, it is also possible to calculate the minimum amplitude. This quantity could be used to determine contours on which the complex polarization does not change, which could potentially be used to automatically identify polarization gradient filaments. The minimum amplitude is attained for an angle $\theta_{\text{min}}$ that satisfies
\begin{align}
\cos 2\theta_{\text{min}} &= \frac{\bigl( \bigl( \frac{\partial Q}{\partial y} \bigr)^2 - \bigl( \frac{\partial Q}{\partial x} \bigr)^2 + \bigl( \frac{\partial U}{\partial y} \bigr)^2 - \bigl( \frac{\partial U}{\partial x} \bigr)^2 \bigr) }{ \sqrt{ \bigl( \bigl( \frac{\partial Q}{\partial x} \bigr)^2 + \bigl( \frac{\partial Q}{\partial y} \bigr)^2 + \bigl( \frac{\partial U}{\partial x} \bigr)^2 + \bigl( \frac{\partial U}{\partial y} \bigr)^2 \bigr)^2 - 4 \bigl( \frac{\partial Q}{\partial x} \frac{\partial U}{\partial y} - \frac{\partial Q}{\partial y} \frac{\partial U}{\partial x} \bigr)^2 } } \text{, and} \label{direc_theta_min_cos}\\
\sin 2\theta_{\text{min}} &= \frac{ - 2 \bigl( \frac{\partial Q}{\partial x} \frac{\partial Q}{\partial y} + \frac{\partial U}{\partial x}  \frac{\partial U}{\partial y} \bigr) }{ \sqrt{ \bigl( \bigl( \frac{\partial Q}{\partial x} \bigr)^2 + \bigl( \frac{\partial Q}{\partial y} \bigr)^2 + \bigl( \frac{\partial U}{\partial x} \bigr)^2 + \bigl( \frac{\partial U}{\partial y} \bigr)^2 \bigr)^2 - 4 \bigl( \frac{\partial Q}{\partial x} \frac{\partial U}{\partial y} - \frac{\partial Q}{\partial y} \frac{\partial U}{\partial x} \bigr)^2 } }  \text{,} \label{direc_theta_min_sin}
\end{align}
and the minimum amplitude of the directional derivative is given by
\begin{align}
\bigg|\frac{\partial \boldsymbol{P}}{\partial s} \bigg|_{\text{min}} &= \Biggl[ \frac{1}{2} \biggl( \biggl(\frac{\partial Q}{\partial x} \biggr)^2 + \biggl(\frac{\partial U}{\partial x} \biggr)^2 + \biggl(\frac{\partial Q}{\partial y} \biggr)^2 + \biggl(\frac{\partial U}{\partial y} \biggr)^2  \biggr) - \nonumber \\ 
 & \frac{1}{2} \sqrt{\biggl( \biggl(\frac{\partial Q}{\partial x} \biggr)^2 + \biggl(\frac{\partial U}{\partial x} \biggr)^2 + \biggl(\frac{\partial Q}{\partial y} \biggr)^2 + \biggl(\frac{\partial U}{\partial y} \biggr)^2 \biggr)^2 - 4 \biggl( \frac{\partial Q}{\partial x} \frac{\partial U}{\partial y} - \frac{\partial Q}{\partial y} \frac{\partial U}{\partial x} \biggr)^2  }  \Biggr]^{1/2}.  \label{direc_min}
\end{align}
We note that the angles $\theta_{\text{max}}$ and $\theta_{\text{min}}$ do not differ by $90^{\circ}$ unless the determinant of the Jacobian of polarization derivatives is zero, and so the directions that maximize and minimize the directional derivative are not necessarily perpendicular.

\subsection{Components of the Directional Derivative} \label{rad_tang_space}
\begin{figure}
\begin{center}
\includegraphics[scale=0.5]{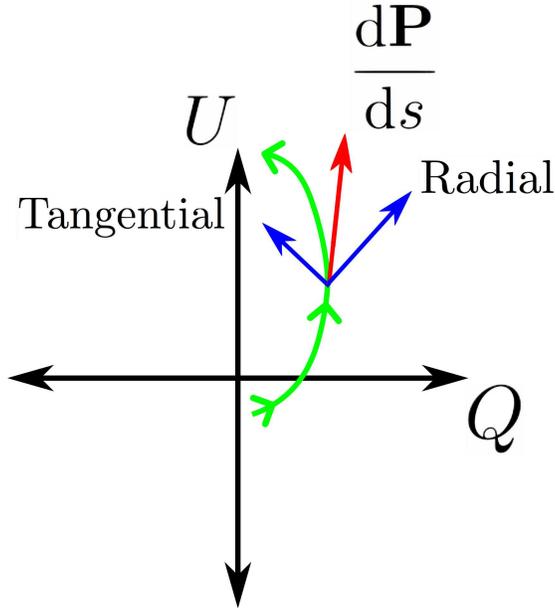}
\caption{A diagram illustrating the calculation of the radial and tangential components of the polarization directional derivative. The polarization directional derivative, expressed as a vector in the \QU plane, is projected onto the radial and azimuthal basis vectors of the \QU plane.}
\label{direc_deriv_comp}
\end{center}
\end{figure}

As the polarization directional derivative is a vector in the \QU plane, it is possible to decompose it into components that are radial and azimuthal in the \QU plane. The radial component quantifies how changes in polarization intensity contribute to the directional derivative, and the azimuthal component, which we will hereafter refer to as the tangential component, quantifies how changes in polarization angle, weighted by polarization intensity, contribute to the directional derivative. This decomposition is illustrated in Fig. \ref{direc_deriv_comp}. The radial and tangential components can provide information on whether individual polarization gradient features are caused by changes in polarization intensity or polarization angle. 

If changes in polarization intensity are dominant for a feature, then this could imply that the amount of depolarization due to the addition of polarization vectors along the line of sight varies significantly between different positions, and it follows that the medium producing the polarized emission may be very turbulent. This is true for both thermal dust emission and for synchrotron emission. If changes in polarization angle are dominant, then this could indicate changes in the regular magnetic field threading the observed region, as this would produce significant changes in the emitted polarization angle in the case of thermal dust emission or synchrotron emission. Additionally, changes in the regular magnetic field may also cause the amount of Faraday rotation along different lines of sight to vary significantly, in the case of synchrotron emission.

To calculate the radial and tangential components of the directional derivative, we consider the radial and azimuthal basis vectors in the \QU plane, which are 
\begin{equation}
\boldsymbol{\hat{r}} = (\cos 2 \psi, \sin 2 \psi) \text{, and } 
\boldsymbol{\hat{\psi}} = (-\sin 2 \psi, \cos 2 \psi)
\end{equation}
respectively. By projecting the directional derivative onto these basis vectors, we obtain the radial and tangential components of the directional derivative, which are
\begin{align}
\frac{\partial \boldsymbol{P}}{\partial s}_{\text{rad}} &= \cos \biggl( \arctan \frac{U}{Q} \biggr) \biggl(\frac{\partial Q}{\partial x} \cos \theta + \frac{\partial Q}{\partial y} \sin \theta \biggr) + \sin \biggl( \arctan \frac{U}{Q} \biggr) \biggl( \frac{\partial U}{\partial x} \cos \theta + \frac{\partial U}{\partial y} \sin \theta \biggr) \text{, and} \label{rad_comp_space} \\
\frac{\partial \boldsymbol{P}}{\partial s}_{\text{tang}} &= -\sin \biggl( \arctan \frac{U}{Q} \biggr) \biggl(\frac{\partial Q}{\partial x} \cos \theta + \frac{\partial Q}{\partial y} \sin \theta \biggr) + \cos \biggl( \arctan \frac{U}{Q} \biggr) \biggl( \frac{\partial U}{\partial x} \cos \theta + \frac{\partial U}{\partial y} \sin \theta \biggr) \label{tang_comp_space}
\end{align}
respectively. As the relationship between the directional derivative and the radial and azimuthal basis vectors changes with translations of the \QU plane, the radial and tangential components of the directional derivative are not translationally invariant quantities, although they are rotationally invariant. However, they are still useful as a means of quantifying the contribution of changes in polarization intensity and polarization angle to the amplitude of the directional derivative. 

As for the directional derivative, it is possible to calculate the value of $\theta$ that maximizes the radial and tangential components of the directional derivative. The maximum of the radial component is
\begin{equation}
\frac{\partial \boldsymbol{P}}{\partial s}_{\text{rad, max}} = \sqrt{ \frac{ \bigl( Q \frac{\partial Q}{\partial x} + U \frac{\partial U}{\partial x} \bigr)^2 + \bigl( Q \frac{\partial Q}{\partial y} + U \frac{\partial U}{\partial y} \bigr)^2 }{ Q^2 + U^2 } }, \label{rad_direc_max}
\end{equation}
and this maximum is obtained for angle $\theta_{\text{rad, max}}$ that satisfies
\begin{align}
\cos \theta_{\text{rad, max}} &= \frac{Q \frac{\partial Q}{\partial x} + U \frac{\partial U}{\partial x}}{ \sqrt{ \bigl( Q  \frac{\partial Q}{\partial x} + U \frac{\partial U}{\partial x}\bigr)^2 + \bigl( Q \frac{\partial Q}{\partial y} + U \frac{\partial U}{\partial y} \bigr)^2 } } \text{, and} \\
\sin \theta_{\text{rad, max}} &= \frac{Q \frac{\partial Q}{\partial y} + U \frac{\partial U}{\partial y}}{ \sqrt{ \bigl( Q  \frac{\partial Q}{\partial x} + U \frac{\partial U}{\partial x}\bigr)^2 + \bigl( Q \frac{\partial Q}{\partial y} + U \frac{\partial U}{\partial y} \bigr)^2 } }.
\end{align}

The maximum value of the tangential component is
\begin{equation}
\frac{\partial \boldsymbol{P}}{\partial s}_{\text{tang, max}} = \sqrt{ \frac{ \bigl( Q \frac{\partial U}{\partial x} - U \frac{\partial Q}{\partial x} \bigr)^2 + \bigl( Q \frac{\partial U}{\partial y} - U \frac{\partial Q}{\partial y} \bigr)^2 }{ Q^2 + U^2 } }, \label{tang_direc_max}
\end{equation}
and this maximum is obtained for angle $\theta_{\text{tang, max}}$, not necessarily orthogonal to $\theta_{\text{rad, max}}$, that satisfies
\begin{align}
\cos \theta_{\text{tang, max}} &= \frac{Q \frac{\partial U}{\partial x} - U \frac{\partial Q}{\partial x}}{ \sqrt{ \bigl( Q  \frac{\partial U}{\partial x} - U \frac{\partial Q}{\partial x}\bigr)^2 + \bigl( Q \frac{\partial U}{\partial y} - U \frac{\partial Q}{\partial y} \bigr)^2 } } \text{, and} \\
\sin \theta_{\text{tang, max}} &= \frac{Q \frac{\partial U}{\partial y} - U \frac{\partial Q}{\partial y}}{ \sqrt{ \bigl( Q  \frac{\partial U}{\partial x} - U \frac{\partial Q}{\partial x}\bigr)^2 + \bigl( Q \frac{\partial U}{\partial y} - U \frac{\partial Q}{\partial y} \bigr)^2 } }. 
\end{align}
By using the relationship between the Stokes parameters and $P$ and $\psi$, we can express the maxima of the radial and tangential components of the directional derivative in terms of the polarization intensity and polarization angle:
\begin{align}
\frac{\partial \boldsymbol{P}}{\partial s}_{\text{rad, max}} &= \sqrt{ \biggl( \frac{\partial P}{\partial x} \biggr)^2 + \biggl( \frac{\partial P}{\partial y} \biggr)^2 }, \label{rad_direc_max_pol} \\
\frac{\partial \boldsymbol{P}}{\partial s}_{\text{tang, max}} &= 2P \sqrt{ \biggl( \frac{\partial \psi}{\partial x} \biggr)^2 + \biggl( \frac{\partial \psi}{\partial y} \biggr)^2 }. \label{tang_direc_max_pol}
\end{align}
Eq. \ref{rad_direc_max_pol} demonstrates that the maximum value of the radial component of the directional derivative is the same as the gradient of polarization intensity\footnote{We note that the gradient of polarization intensity is different to the amplitude of the polarization gradient.}, and Eq. \ref{tang_direc_max_pol} shows that the maximum of the tangential component is the same as the gradient of the polarization angle, weighted by polarization intensity. The advantage of Eqs. \ref{rad_direc_max} and \ref{tang_direc_max} is that they provide a simple means of comparing the importance of changes in polarization intensity and polarization angle throughout an image, and allow us to calculate these gradients without calculating the derivatives of the polarization intensity or polarization angle, which suffer from noise bias. However, Eqs. \ref{rad_direc_max} and \ref{tang_direc_max} involve square roots of squared quantities, and so are likely subject to a noise bias that is similar to the noise bias that affects polarization intensity \citep{Wardle1974, Simmons1985}. A statistical analysis is required to determine whether Eqs. \ref{rad_direc_max} and \ref{tang_direc_max} provide a more robust method of calculating gradients of the polarization intensity or polarization angle. This analysis could be performed by calculating the gradients of the polarization intensity and polarization angle for mock polarization images produced from simulations of turbulence, and comparing how the methods of calculating the gradients behave when noise is added to the image. Such an analysis is well suited to a paper that investigates how the diagnostics derived in this paper behave in the presence of noise, and is beyond the scope of this paper.

From Eqs. \ref{rad_direc_max} and \ref{tang_direc_max}, we find that the maxima of the radial and tangential components of the directional derivative are related to the polarization gradient according to
\begin{equation}
|\nabla {\boldsymbol{P}}|^2 = \biggl( \frac{\partial \boldsymbol{P}}{\partial s}_{\text{rad, max}} \biggr)^2 + \biggl( \frac{\partial \boldsymbol{P}}{\partial s}_{\text{tang, max}} \biggr)^2. \label{grad_comp_rel}
\end{equation}
The polarization gradient is hence equal to the quadrature of the maximum values of the radial and tangential components of the directional derivative, and hence to the quadrature of the gradients of polarization intensity and polarization angle. While Eq. \ref{grad_comp_rel} suggests a Pythagorean relationship between these three quantities, we note that the radial and tangential components of the directional derivative are maximized in directions that are not perpendicular, in general. We also note that the radial and tangential components are not necessarily maximized for the same value of $\theta$, and so the radial and tangential components of the polarization gradient are not necessarily equal to the maximal values of these components.

\section{Second Order Spatial Derivatives and Curvature}
\label{2Space}
In the previous section we derived a formula for the speed at which we traverse a path in the \QU plane as we move across an image, as this quantity is a property of the traversed path that is invariant under rotations and translations of the \QU plane. Another invariant property of the traversed path is its curvature, which we derive in this section. As the curvature is independent of the speed at which we traverse the path, the curvature has the potential to provide a new way of visualizing magnetoionic turbulence, that is complementary to the gradient.

To derive the curvature of a path in the \QU plane, we first need to re-parametrize it by the arc length of the path, to ensure that the path is traversed at unit speed. This step is important in defining a unique value of the curvature of the path, as traversing the path at different speeds will lead to different measured accelerations (see \citealt{DoCarmo2016} for more information).

We let $z(s)$ be the arc length in the \QU plane, such that $z=0$ when $s=0$,
\begin{equation}
z(s) = \int_0^s \bigg| \frac{\partial \boldsymbol{P}}{\partial s'} \bigg| \, \mathrm{d}s'. \label{arc_length}
\end{equation}
We now re-parametrize the complex polarization vector to be a function of arc length, $\boldsymbol{P}(z)$. We note that we can only make this re-parametrization if the directional derivative is non-zero in the direction that we wish to calculate the curvature in. We can then calculate the unit-length velocity vector of the re-parametrized curve,
\begin{equation}
\frac{\partial \boldsymbol{P}}{\partial z} = \biggl( \frac{\partial Q}{\partial z}, \frac{\partial U}{\partial z} \biggr),
\end{equation}
and the acceleration vector, which is perpendicular to the velocity vector, is
\begin{equation}
\frac{\partial^2 \boldsymbol{P}}{\partial z^2} = \biggl( \frac{\partial^2 Q}{\partial z^2}, \frac{\partial^2 U}{\partial z^2} \biggr). \label{direc_curv_vec}
\end{equation}
The amplitude of the acceleration vector gives the curvature of the path in the \QU plane. Although we omit the derivation here, it is possible to calculate the second derivatives of the Stokes parameters with respect to $z$, in terms of $x$, $y$ and $\theta$, by using the chain rule and the Fundamental Theorem of Calculus. We call the amplitude of the acceleration vector the `polarization directional curvature', $k_s(x,y,\lambda^2;\theta)$:
\begin{align}
k_s(x,y,\lambda^2;\theta) &=  \bigg| \frac{\partial \boldsymbol{P}}{\partial s} \bigg|^{-3} \biggl[ \cos^3 \theta \biggl( \frac{\partial Q}{\partial x} \frac{\partial^2 U}{\partial x^2} - \frac{\partial U}{\partial x} \frac{\partial^2 Q}{\partial x^2} \biggr) + 2 \cos^2 \theta \sin \theta \biggl( \frac{\partial Q}{\partial x} \frac{\partial^2 U}{\partial x \partial y} - \frac{\partial U}{\partial x} \frac{\partial^2 Q}{\partial x \partial y} \biggr) + \nonumber \\
&  \cos^2 \theta \sin \theta \biggl( \frac{\partial Q}{\partial y} \frac{\partial^2 U}{\partial x^2} - \frac{\partial U}{\partial y} \frac{\partial^2 Q}{\partial x^2} \biggr) + 2 \cos \theta \sin^2 \theta \biggl( \frac{\partial Q}{\partial y} \frac{\partial^2 U}{\partial x \partial y} - \frac{\partial U}{\partial y} \frac{\partial^2 Q}{\partial x \partial y} \biggr) + \nonumber \\
&  \cos \theta \sin^2 \theta \biggl( \frac{\partial Q}{\partial x} \frac{\partial^2 U}{\partial y^2} - \frac{\partial U}{\partial x} \frac{\partial^2 Q}{\partial y^2} \biggr) + \sin^3 \theta \biggl( \frac{\partial Q}{\partial y} \frac{\partial^2 U}{\partial y^2} - \frac{\partial U}{\partial y} \frac{\partial^2 Q}{\partial y^2} \biggr) \biggr], \label{pol_direc_curv}
\end{align}
and we find that this quantity is rotationally and translationally invariant in the \QU plane by using the method outlined in Section \ref{framework}. Note that in Eq. \ref{pol_direc_curv}, we divide by the cube of the amplitude of the directional derivative, which reflects the fact that it is not valid to re-parametrize by arc length if the directional derivative is zero at a particular point in an image, in a specified direction. A convenient way of avoiding the problem of zero directional derivative is to calculate the directional curvature in the direction that maximizes the directional derivative, $\theta_{\text{max}}$. This ensures that the curvature is calculated in a direction for which it is valid to calculate the curvature, at every pixel of an image. The only pixels that are exceptions to this are those for which the maximum amplitude of the directional derivative is zero, and in this case the curvature cannot be calculated in any direction for the pixel.

We also note that because the directional derivative (which is approximately proportional to polarization intensity) is cubed in the denominator of Eq. \ref{pol_direc_curv}, and the numerator of Eq. \ref{pol_direc_curv} scales as the polarization intensity squared, the directional curvature is approximately inversely proportional to polarization intensity. This means that the curvature can be very large in areas of high noise, and these areas should be masked prior to calculating the curvature. 

Unlike the directional derivative, the polarization directional curvature can be either positive or negative. Positive curvature corresponds to a path that we traverse in an anti-clockwise direction (the cross product of the unit velocity vector and the acceleration vector points out of the \QU plane), and negative curvature corresponds to a path that is traversed in a clockwise direction. Reversing the direction in which we calculate the directional curvature in the image plane causes the direction that we traverse the path in the \QU plane to reverse, and so it causes the sign of the curvature to flip. This means that we only need to calculate the directional curvature for values of $\theta$ between $0^{\circ}$ and $180^{\circ}$, as the curvature for other values of $\theta$ will only differ in sign from the curvature for values of $\theta$ in this range.

While it should be possible to determine an analytic expression for the value of $\theta$ that maximizes the directional curvature, and the maximum directional curvature, it is mathematically arduous to do so. Instead, the maximum value of the directional curvature can be determined numerically, by calculating the curvature for various values of $\theta$.

Any path in a two-dimensional plane, such as the \QU plane, is completely defined by its velocity vector and its curvature, and hence we do not expect there to be any other rotationally and translationally invariant quantities that are independent of the directional derivative and directional curvature, and which only involve spatial derivatives of the complex polarization, in a specified direction of the image.

\section{First Order Wavelength Derivatives}
\label{1Wave}
In this section, we consider how the complex polarization vector changes as we vary the observing wavelength, at a given pixel of the image, as shown in Fig. \ref{wav_deriv_path}. For the case of synchrotron emission from a discrete or diffuse source, these wavelength derivatives have the potential to reveal the statistics of the turbulent Faraday rotating medium, as they probe the interference of polarization subjected to differing amounts of Faraday rotation, similar to the Polarization Frequency Analysis method developed by \cite{Lazarian2016}.

We define the polarization wavelength derivative as a vector in the \QU plane by Eq. \ref{pol_velocity_wav}. As was the case for the polarization directional derivative, the amplitude of this vector is rotationally and translationally invariant in the \QU plane, and it is given by
\begin{equation}
\bigg| \frac{\partial \boldsymbol{P}}{\partial \lambda^2} \bigg| = \sqrt{\biggl( \frac{\partial Q}{\partial \lambda^2} \biggr)^2 + \biggl( \frac{\partial U}{\partial \lambda^2} \biggr)^2}. \label{pol_wav_deriv}
\end{equation}
We will refer to Eq. \ref{pol_wav_deriv} as the `polarization wavelength derivative'. We note that \cite{Lazarian2016} investigated the correlation function of the vector form of the polarization wavelength derivative, and found that it was sensitive to Faraday rotation. It is possible that the correlation function of Eq. \ref{pol_wav_deriv} is also sensitive to Faraday rotation, whilst also being rotationally and translationally invariant in the \QU plane.

As was done for the directional derivative (Eq. \ref{direc_deriv}), it is possible to calculate the radial and tangential components of the wavelength derivative in the \QU plane. The radial and tangential components of the wavelength derivative are given by
\begin{align}
\frac{\partial \boldsymbol{P}}{\partial \lambda^2}_{\text{rad}} &= \cos \biggl( \arctan \frac{U}{Q} \biggr) \frac{\partial Q}{\partial \lambda^2} + \sin \biggl( \arctan \frac{U}{Q} \biggr) \frac{\partial U}{\partial \lambda^2} \text{, and} \label{rad_comp_wav} \\
\frac{\partial \boldsymbol{P}}{\partial \lambda^2}_{\text{tang}} &= -\sin \biggl( \arctan \frac{U}{Q} \biggr) \frac{\partial Q}{\partial \lambda^2} + \cos \biggl( \arctan \frac{U}{Q} \biggr) \frac{\partial U}{\partial \lambda^2}, \label{tang_comp_wav}
\end{align}
respectively. The radial and tangential components of the wavelength derivative are not translationally invariant in the \QU plane, but are rotationally invariant. The radial component of the wavelength derivative quantifies how changes in polarization intensity with wavelength contribute to the wavelength derivative, and the tangential component quantifies how changes in the polarization angle, weighted by the polarization intensity, contribute to the wavelength derivative. As was the case with the radial and tangential components of the directional derivative, if the wavelength derivative of synchrotron emission is dominated by changes in polarization intensity, it implies that the emitting medium is highly turbulent, due to depolarization of Faraday rotated polarization vectors along the line of sight. If the wavelength derivative of synchrotron emission is dominated by changes in polarization angle, it may indicate the presence of a strong, regular magnetic field that is causing significant Faraday rotation of the polarization vectors.

\section{Second Order Wavelength Derivatives and Curvature}
\label{2Wave}
\begin{figure*}
\begin{center}
\includegraphics[scale=0.5]{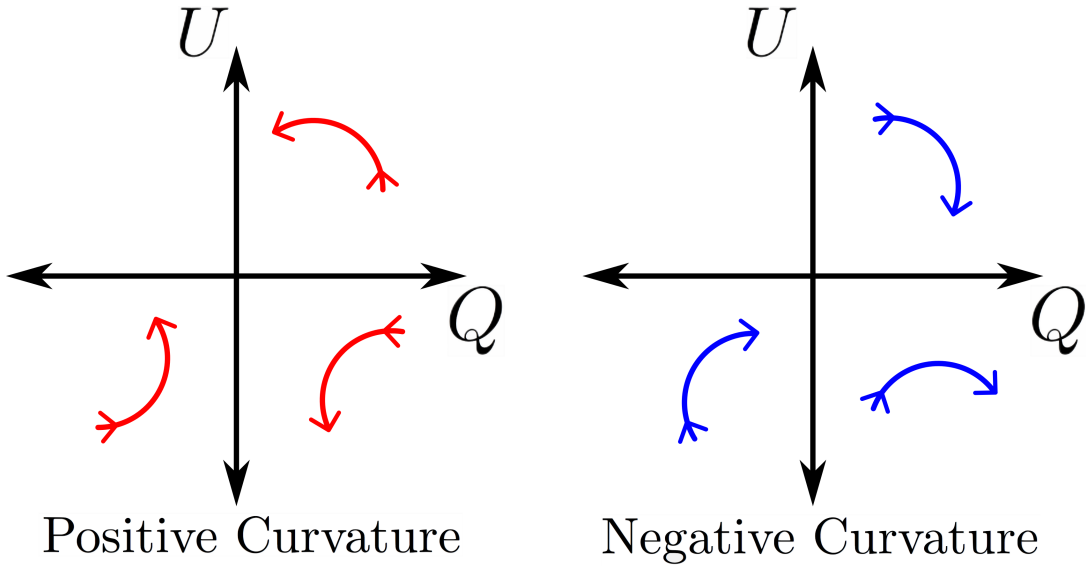}
\caption{A diagram illustrating the interpretation of positive (anti-clockwise, left) and negative (clockwise, right) curvature. All curves on the left are the same, as are all curves on the right. Note that the rotation measure, defined as the rate of change of polarization angle, is different for all of the curves on the left. However, the polarization wavelength derivative and wavelength curvature are the same for all of these curves.}
\label{wav_curv_examples}
\end{center}
\end{figure*}

We can also calculate the curvature of the path traversed in the \QU plane as the wavelength changes, by first defining the arc length of this path, $l(\lambda^2)$, by
\begin{equation}
l(\lambda^2) = \int_{\lambda^2_{\text{min}}}^{\lambda^2} \bigg| \frac{\partial \boldsymbol{P}}{\partial \lambda'^2} \bigg| \, \mathrm{d} \lambda'^2,
\end{equation}
where $\lambda^2_{\text{min}}$ is calculated at the smallest observed wavelength. Assuming that the wavelength derivative is non-zero at this pixel, and at this wavelength, we can re-parametrize the path in the \QU plane as $\boldsymbol{P}(l)$. Then the first order derivative of $\boldsymbol{P}(l)$ with respect to $l$ provides the unit-length velocity along the path, and the second order derivative 
\begin{equation}
\frac{\partial^2 \boldsymbol{P}}{\partial l^2} = \biggl( \frac{\partial^2 Q}{\partial l^2}, \frac{\partial^2 U}{\partial l^2}  \biggr) \label{wav_curv_vec}
\end{equation}
is the acceleration vector, whose amplitude gives the curvature of the path. We call this the `polarization wavelength curvature',
\begin{equation}
k_{\lambda}(x,y,\lambda^2) = \bigg| \frac{\partial \boldsymbol{P}}{\partial \lambda^2} \bigg|^{-3} \biggl[ \frac{\partial Q}{\partial \lambda^2} \frac{\partial^2 U}{\partial (\lambda^2)^2} - \frac{\partial U}{\partial \lambda^2} \frac{\partial^2 Q}{\partial (\lambda^2)^2} \biggr],
\end{equation} 
and it is rotationally and translationally invariant in the \QU plane. Similar to the directional curvature, the wavelength curvature is approximately inversely proportional to polarization intensity, and can be positive or negative, where positive curvature corresponds to a path that is traversed anti-clockwise, and negative curvature corresponds to a path that is traversed clockwise, as depicted in Fig. \ref{wav_curv_examples}. In Fig. \ref{wav_curv_examples} we display paths in the \QU plane that have the same positive curvature (on the left) and the same negative curvature (on the right), but that have been rotated and translated around the \QU plane. We note that the rotation measure, defined as the rate of change of the polarization angle, and commonly used to measure Faraday rotation due to a magnetoionic medium, is different for all of the paths on the left of Fig. \ref{wav_curv_examples}. The polarization wavelength derivative and wavelength curvature, on the other hand, are invariant, and so may provide a more robust means of studying Faraday rotation.

\section{Discussion}
\label{discuss}
The polarization diagnostics that we have presented have potential applications to different forms of polarized emission. For observations of polarized, thermal dust emission, and diffuse synchrotron emission, the polarization directional derivative and directional curvature describe spatial variations in the magnetic field and density, which may be caused by vorticity, shear, or shocks. For these different emission types, interstellar turbulence plays a large role, and so statistics of the directional derivative and directional curvature could be used to constrain properties of turbulence in different phases of the ISM, by using the same approach as \cite{Burkhart2012}. 

For polarized dust emission, the polarization wavelength derivative may trace how the density and magnetic field of dust at different temperatures changes, as cooler dust may dominate the observed polarized emission at long wavelengths. When the polarization wavelength derivative is applied to radio-frequency synchrotron emission, it should mostly probe Faraday rotation within and in front of the emitting region (this is the dominant wavelength-dependent mechanism), providing insight on the turbulent magnetoionic medium along the line of sight. In both cases, statistics of polarization wavelength derivative and wavelength curvature maps could be used to help constrain properties of turbulence. 

Polarization wavelength derivatives can also be applied to discrete sources of polarized synchrotron emission, such as unresolved radio galaxies. In this context, the wavelength derivative and wavelength curvature can provide a new way of studying the interference of different polarized components within the telescope beam, such as the lobes and nucleus of a radio galaxy. In particular, as the wavelength derivative and wavelength curvature completely describe the dependence of polarization on wavelength in a manner that is robust to observational artefacts, they could be incorporated into the $QU$-fitting method that is currently used to fit models of polarized emission from radio galaxies to observations (e.g. \citealt{OSullivan2012}, \citealt{Anderson2016}). 

Although we do not investigate the noise statistics of our derived diagnostics in this paper, we note that because all of our invariant diagnostics involve a square root of squared quantities (the polarization directional curvature and polarization wavelength curvature have such terms in their denominators), they will be subject to a noise bias similar to the Ricean bias that affects polarization intensity. None of our diagnostics have a Ricean noise distribution, and so it will be necessary to analyze the noise properties of each diagnostic in order to understand the behaviour of these diagnostics when applied to a noisy signal.

\section{Conclusions}
\label{concl}
We have derived new quantities that are rotationally and translationally invariant in the Q-U plane, that can be applied to polarimetric observations of any object, at any wavelength. These include the polarization directional derivative and directional curvature, which completely describe spatial changes of the complex polarization in a specified direction. As the generalized polarization gradient is very similar to the polarization gradient, it traces spatial variations in magnetoionic media, such as vorticity, shear, and shocks. We have also derived the polarization wavelength derivative and wavelength curvature, which completely describe spectral changes of the complex polarization. These two quantities should provide a robust, sensitive method of studying Faraday rotation. In Paper II, we will examine these diagnostics for the case of diffuse, radio-frequency synchrotron emission arising from the turbulent, magnetized, warm-ionized medium.

\section*{Acknowledgements}

C.~A.~H. acknowledges financial support received via an Australian Postgraduate Award, and a Vice Chancellor's Research Scholarship awarded by the University of Sydney. B.~M.~G. acknowledges the support of the Natural Sciences and Engineering Research Council of Canada (NSERC) through grant RGPIN-2015-05948 and of a Canada Research Chair. N.~M.~M.-G. acknowledges the support of the Australian Research Council through grant FT150100024. The Dunlap Institute for Astronomy and Astrophysics is funded through an endowment established by the David Dunlap family and the University of Toronto. This research made use of Astropy, a community-developed core Python package for Astronomy \citep{Astropy2013}.

\appendix
\section{Appendix A: Mixed Derivatives}
\label{Mixed}
In addition to studying how the complex polarization varies across the image and with wavelength, it is possible to examine how the complex polarization depends on spatial and spectral changes simultaneously. The resultant quantity can be considered as the spatial derivative of the wavelength derivative, or equivalently as the wavelength derivative of the directional derivative. We obtain this quantity by differentiating Eq. \ref{pol_velocity_space} with respect to wavelength
\begin{equation}
\frac{\partial}{\partial \lambda^2} \biggl( \frac{\partial \boldsymbol{P}}{\partial s} \biggr) = \biggl( \cos \theta \frac{\partial^2 Q}{\partial \lambda^2 \partial x} + \sin \theta \frac{\partial^2 Q}{\partial \lambda^2 \partial y} , \cos \theta \frac{\partial^2 U}{\partial \lambda^2 \partial x} + \sin \theta \frac{\partial^2 U}{\partial \lambda^2 \partial y}  \biggr). \label{mixed_deriv_vec}
\end{equation}
Eq. \ref{mixed_deriv_vec} quantifies how spatial changes in polarization depend on wavelength, and the amplitude of this vector, which we call the `polarization mixed derivative',
\begin{align}
\bigg| \frac{\partial}{\partial \lambda^2} \biggl( \frac{\partial \boldsymbol{P}}{\partial s} \biggr) \bigg| &= \biggl[ \cos^2 \theta \biggl( \biggl( \frac{\partial^2 Q}{\partial \lambda^2 \partial x} \biggr)^2 + \biggl( \frac{\partial^2 U}{\partial \lambda^2 \partial x} \biggr)^2  \biggr) + 2 \sin \theta \cos \theta \biggl( \frac{\partial^2 Q}{\partial \lambda^2 \partial x} \frac{\partial^2 Q}{\partial \lambda^2 \partial y} + \frac{\partial^2 U}{\partial \lambda^2 \partial x} \frac{\partial^2 U}{\partial \lambda^2 \partial y} \biggr) + \nonumber \\
& \qquad \quad \sin^2 \theta \biggl( \biggl( \frac{\partial^2 Q}{\partial \lambda^2 \partial y} \biggr)^2 + \biggl( \frac{\partial^2 U}{\partial \lambda^2 \partial y} \biggr)^2 \biggr) \biggr]^{1/2}, \label{mix_deriv_amp}
\end{align}
is rotationally and translationally invariant in the \QU plane. As was the case with the directional derivative, we can maximize Eq. \ref{mix_deriv_amp} with respect to $\theta$. We find that the maximal mixed derivative is given by
\begin{align}
\bigg| \frac{\partial}{\partial \lambda^2} \biggl( \frac{\partial \boldsymbol{P}}{\partial s} \biggr) \bigg|_{\text{max}} &= \biggl[ \frac{1}{2} \biggl( \biggl( \frac{\partial^2 Q}{\partial \lambda^2 \partial x} \biggr)^2 + \biggl( \frac{\partial^2 U}{\partial \lambda^2 \partial x} \biggr)^2 + \biggl( \frac{\partial^2 Q}{\partial \lambda^2 \partial y} \biggr)^2 + \biggl( \frac{\partial^2 U}{\partial \lambda^2 \partial y} \biggr)^2 \biggr) + \nonumber \\
& \frac{1}{2} \sqrt{ \biggl( \biggl( \frac{\partial^2 Q}{\partial \lambda^2 \partial x} \biggr)^2 + \biggl( \frac{\partial^2 U}{\partial \lambda^2 \partial x} \biggr)^2 + \biggl( \frac{\partial^2 Q}{\partial \lambda^2 \partial y} \biggr)^2 + \biggl( \frac{\partial^2 U}{\partial \lambda^2 \partial y} \biggr)^2 \biggr)^2 - 4 \biggl( \frac{\partial^2 Q}{\partial \lambda^2 \partial x} \frac{\partial^2 U}{\partial \lambda^2 \partial y} - \frac{\partial^2 Q}{\partial \lambda^2 \partial y} \frac{\partial^2 U}{\partial \lambda^2 \partial x} \biggr)^2 } \biggr]^{1/2}, \label{mix_deriv_amp_max}
\end{align}
and this maximal value is obtained for angle $\theta_{\text{mix, max}}$ that satisfies both of
\begin{align}
\cos 2\theta_{\text{mix, max}} &= \frac{ - \bigl( \bigl( \frac{\partial^2 Q}{\partial \lambda^2 \partial y} \bigr)^2 - \bigl( \frac{\partial^2 Q}{\partial \lambda^2 \partial x} \bigr)^2 + \bigl( \frac{\partial^2 U}{\partial \lambda^2 \partial y} \bigr)^2 - \bigl( \frac{\partial^2 U}{\partial \lambda^2 \partial x} \bigr)^2 \bigr) }{ \sqrt{ \bigl( \bigl( \frac{\partial^2 Q}{\partial \lambda^2 \partial x} \bigr)^2 + \bigl( \frac{\partial^2 Q}{\partial \lambda^2 \partial y} \bigr)^2 + \bigl( \frac{\partial^2 U}{\partial \lambda^2 \partial x} \bigr)^2 + \bigl( \frac{\partial^2 U}{\partial \lambda^2 \partial y} \bigr)^2 \bigr)^2 - 4 \bigl( \frac{\partial^2 Q}{\partial \lambda^2 \partial x} \frac{\partial^2 U}{\partial \lambda^2 \partial y} - \frac{\partial^2 Q}{\partial \lambda^2 \partial y} \frac{\partial^2 U}{\partial \lambda^2 \partial x} \bigr)^2 } } \text{, and} \label{mix_theta_max_cos}\\
\sin 2\theta_{\text{mix, max}} &= \frac{ 2 \bigl( \frac{\partial^2 Q}{\partial \lambda^2 \partial x} \frac{\partial^2 Q}{\partial \lambda^2 \partial y} + \frac{\partial^2 U}{\partial \lambda^2 \partial x}  \frac{\partial^2 U}{\partial \lambda^2 \partial y} \bigr) }{ \sqrt{ \bigl( \bigl( \frac{\partial^2 Q}{\partial \lambda^2 \partial x} \bigr)^2 + \bigl( \frac{\partial^2 Q}{\partial \lambda^2 \partial y} \bigr)^2 + \bigl( \frac{\partial^2 U}{\partial \lambda^2 \partial x} \bigr)^2 + \bigl( \frac{\partial^2 U}{\partial \lambda^2 \partial y} \bigr)^2 \bigr)^2 - 4 \bigl( \frac{\partial^2 Q}{\partial \lambda^2 \partial x} \frac{\partial^2 U}{\partial \lambda^2 \partial y} - \frac{\partial^2 Q}{\partial \lambda^2 \partial y} \frac{\partial^2 U}{\partial \lambda^2 \partial x} \bigr)^2 } }  \text{.} \label{mix_theta_max_sin}
\end{align}

Another diagnostic that combines spatial and spectral derivatives is the angle between the directional derivative (Eq. \ref{pol_velocity_space}) and wavelength derivative (Eq. \ref{pol_velocity_wav}) in the \QU plane. This angle is rotationally and translationally invariant in the \QU plane, and quantifies how different the spatial and spectral changes in the polarization are. Similarly, the angle between the directional curvature (Eq. \ref{direc_curv_vec}) and the wavelength curvature (Eq. \ref{wav_curv_vec}) is rotationally and translationally invariant in the \QU plane.

\bibliography{ref_list}

\begin{thebibliography}{}
\expandafter\ifx\csname natexlab\endcsname\relax\def\natexlab#1{#1}\fi

\bibitem[{{Anderson} {et~al.}(2016){Anderson}, {Gaensler}, \&
  {Feain}}]{Anderson2016}
{Anderson}, C.~S., {Gaensler}, B.~M., \& {Feain}, I.~J. 2016, \apj, 825, 59

\bibitem[{{Astropy Collaboration} {et~al.}(2013){Astropy Collaboration},
  {Robitaille}, {Tollerud}, {Greenfield}, {Droettboom}, {Bray}, {Aldcroft},
  {Davis}, {Ginsburg}, {Price-Whelan}, {Kerzendorf}, {Conley}, {Crighton},
  {Barbary}, {Muna}, {Ferguson}, {Grollier}, {Parikh}, {Nair}, {Unther},
  {Deil}, {Woillez}, {Conseil}, {Kramer}, {Turner}, {Singer}, {Fox}, {Weaver},
  {Zabalza}, {Edwards}, {Azalee Bostroem}, {Burke}, {Casey}, {Crawford},
  {Dencheva}, {Ely}, {Jenness}, {Labrie}, {Lim}, {Pierfederici}, {Pontzen},
  {Ptak}, {Refsdal}, {Servillat}, \& {Streicher}}]{Astropy2013}
{Astropy Collaboration}, {Robitaille}, T.~P., {Tollerud}, E.~J., {et~al.} 2013,
  \aap, 558, A33

\bibitem[{{Basu} {et~al.}(2017){Basu}, {Mao}, {Kepley}, {Robishaw}, {Zweibel},
  \& {Gallagher}}]{Basu2017}
{Basu}, A., {Mao}, S.~A., {Kepley}, A.~A., {et~al.} 2017, \mnras, 464, 1003

\bibitem[{{Burkhart} {et~al.}(2012){Burkhart}, {Lazarian}, \&
  {Gaensler}}]{Burkhart2012}
{Burkhart}, B., {Lazarian}, A., \& {Gaensler}, B.~M. 2012, \apj, 749, 145

\bibitem[{{Caldwell} {et~al.}(2017){Caldwell}, {Hirata}, \&
  {Kamionkowski}}]{Caldwell2016}
{Caldwell}, R.~R., {Hirata}, C., \& {Kamionkowski}, M. 2017, \apj, 839, 91

\bibitem[{{Carretti}(2010)}]{Carretti2010}
{Carretti}, E. 2010, in Astronomical Society of the Pacific Conference Series,
  Vol. 438, The Dynamic Interstellar Medium: A Celebration of the Canadian
  Galactic Plane Survey, ed. R.~{Kothes}, T.~L. {Landecker}, \& A.~G. {Willis},
  276

\bibitem[{{Carretti} {et~al.}(2013){Carretti}, {Crocker}, {Staveley-Smith},
  {Haverkorn}, {Purcell}, {Gaensler}, {Bernardi}, {Kesteven}, \&
  {Poppi}}]{Carretti2013}
{Carretti}, E., {Crocker}, R.~M., {Staveley-Smith}, L., {et~al.} 2013, \nat,
  493, 66

\bibitem[{{Crutcher} {et~al.}(2010){Crutcher}, {Wandelt}, {Heiles},
  {Falgarone}, \& {Troland}}]{Crutcher2010}
{Crutcher}, R.~M., {Wandelt}, B., {Heiles}, C., {Falgarone}, E., \& {Troland},
  T.~H. 2010, \apj, 725, 466

\bibitem[{do~Carmo(2016)}]{DoCarmo2016}
do~Carmo, M. 2016, Differential Geometry of Curves and Surfaces: Revised and
  Updated Second Edition, Dover Books on Mathematics (Dover Publications)

\bibitem[{{Fletcher} {et~al.}(2011){Fletcher}, {Beck}, {Shukurov},
  {Berkhuijsen}, \& {Horellou}}]{Fletcher2011}
{Fletcher}, A., {Beck}, R., {Shukurov}, A., {Berkhuijsen}, E.~M., \&
  {Horellou}, C. 2011, \mnras, 412, 2396

\bibitem[{{Gaensler} {et~al.}(2015){Gaensler}, {Agudo}, {Akahori}, {Banfield},
  {Beck}, {Carretti}, {Farnes}, {Haverkorn}, {Heald}, {Jones}, {Landecker},
  {Mao}, {Norris}, {O'Sullivan}, {Rudnick}, {Schnitzeler}, {Seymour}, \&
  {Sun}}]{Gaensler2015}
{Gaensler}, B., {Agudo}, I., {Akahori}, T., {et~al.} 2015, Advancing
  Astrophysics with the Square Kilometre Array (AASKA14), 103

\bibitem[{{Gaensler} {et~al.}(2011){Gaensler}, {Haverkorn}, {Burkhart},
  {Newton-McGee}, {Ekers}, {Lazarian}, {McClure-Griffiths}, {Robishaw},
  {Dickey}, \& {Green}}]{Gaensler2011}
{Gaensler}, B.~M., {Haverkorn}, M., {Burkhart}, B., {et~al.} 2011, \nat, 478,
  214

\bibitem[{{Gardner} \& {Whiteoak}(1966)}]{Gardner1966}
{Gardner}, F.~F., \& {Whiteoak}, J.~B. 1966, \araa, 4, 245

\bibitem[{{Hamaker} \& {Bregman}(1996)}]{Hamaker1996}
{Hamaker}, J.~P., \& {Bregman}, J.~D. 1996, \aaps, 117, 161

\bibitem[{{Herron} {et~al.}(2017){Herron}, {Geisbuesch}, {Landecker}, {Kothes},
  {Gaensler}, {Lewis}, {McClure-Griffiths}, \& {Petroff}}]{Herron2017}
{Herron}, C.~A., {Geisbuesch}, J., {Landecker}, T.~L., {et~al.} 2017, \apj,
  835, 210

\bibitem[{{Hill} {et~al.}(2017){Hill}, {Landecker}, {Carretti}, {Douglas},
  {Sun}, {Gaensler}, {Mao}, {McClure-Griffiths}, {Reich}, {Wolleben}, {Dickey},
  {Gray}, {Haverkorn}, {Leahy}, \& {Schnitzeler}}]{Hill2017}
{Hill}, A.~S., {Landecker}, T.~L., {Carretti}, E., {et~al.} 2017, \mnras, 467,
  4631

\bibitem[{{Iacobelli} {et~al.}(2014){Iacobelli}, {Burkhart}, {Haverkorn},
  {Lazarian}, {Carretti}, {Staveley-Smith}, {Gaensler}, {Bernardi}, {Kesteven},
  \& {Poppi}}]{Iacobelli2014}
{Iacobelli}, M., {Burkhart}, B., {Haverkorn}, M., {et~al.} 2014, \aap, 566, A5

\bibitem[{{IAU}(1974)}]{IAU1974}
{IAU}. 1974, {Transactions of the IAU Vol. 15B (1973) 166}

\bibitem[{{Johnston-Hollitt} {et~al.}(2015){Johnston-Hollitt}, {Govoni},
  {Beck}, {Dehghan}, {Pratley}, {Akahori}, {Heald}, {Agudo}, {Bonafede},
  {Carretti}, {Clarke}, {Colafrancesco}, {Ensslin}, {Feretti}, {Gaensler},
  {Haverkorn}, {Mao}, {Oppermann}, {Rudnick}, {Scaife}, {Schnitzeler}, {Stil},
  {Taylor}, \& {Vacca}}]{Johnston2015}
{Johnston-Hollitt}, M., {Govoni}, F., {Beck}, R., {et~al.} 2015, Advancing
  Astrophysics with the Square Kilometre Array (AASKA14), 92

\bibitem[{{Kaczmarek} {et~al.}(2017){Kaczmarek}, {Purcell}, {Gaensler},
  {McClure-Griffiths}, \& {Stevens}}]{Kaczmarek2017}
{Kaczmarek}, J.~F., {Purcell}, C.~R., {Gaensler}, B.~M., {McClure-Griffiths},
  N.~M., \& {Stevens}, J. 2017, \mnras, 467, 1776

\bibitem[{{Landecker} {et~al.}(2010){Landecker}, {Reich}, {Reid}, {Reich},
  {Wolleben}, {Kothes}, {Uyan{\i}ker}, {Gray}, {Del Rizzo}, {F{\"u}rst},
  {Taylor}, \& {Wielebinski}}]{Landecker2010}
{Landecker}, T.~L., {Reich}, W., {Reid}, R.~I., {et~al.} 2010, \aap, 520, A80

\bibitem[{{Lazarian} \& {Pogosyan}(2016)}]{Lazarian2016}
{Lazarian}, A., \& {Pogosyan}, D. 2016, \apj, 818, 178

\bibitem[{{Lee} {et~al.}(2016){Lee}, {Lazarian}, \& {Cho}}]{Lee2016}
{Lee}, H., {Lazarian}, A., \& {Cho}, J. 2016, \apj, 831, 77

\bibitem[{{Lenc} {et~al.}(2016){Lenc}, {Gaensler}, {Sun}, {Sadler}, {Willis},
  {Barry}, {Beardsley}, {Bell}, {Bernardi}, {Bowman}, {Briggs}, {Callingham},
  {Cappallo}, {Carroll}, {Corey}, {de Oliveira-Costa}, {Deshpande}, {Dillon},
  {Dwarkanath}, {Emrich}, {Ewall-Wice}, {Feng}, {For}, {Goeke}, {Greenhill},
  {Hancock}, {Hazelton}, {Hewitt}, {Hindson}, {Hurley-Walker},
  {Johnston-Hollitt}, {Jacobs}, {Kapi{\'n}ska}, {Kaplan}, {Kasper}, {Kim},
  {Kratzenberg}, {Line}, {Loeb}, {Lonsdale}, {Lynch}, {McKinley}, {McWhirter},
  {Mitchell}, {Morales}, {Morgan}, {Morgan}, {Murphy}, {Neben}, {Oberoi},
  {Offringa}, {Ord}, {Paul}, {Pindor}, {Pober}, {Prabu}, {Procopio}, {Riding},
  {Rogers}, {Roshi}, {Udaya Shankar}, {Sethi}, {Srivani}, {Staveley-Smith},
  {Subrahmanyan}, {Sullivan}, {Tegmark}, {Thyagarajan}, {Tingay}, {Trott},
  {Waterson}, {Wayth}, {Webster}, {Whitney}, {Williams}, {Williams}, {Wu},
  {Wyithe}, \& {Zheng}}]{Lenc2016}
{Lenc}, E., {Gaensler}, B.~M., {Sun}, X.~H., {et~al.} 2016, \apj, 830, 38

\bibitem[{{Mao} {et~al.}(2015){Mao}, {Zweibel}, {Fletcher}, {Ott}, \&
  {Tabatabaei}}]{Mao2015}
{Mao}, S.~A., {Zweibel}, E., {Fletcher}, A., {Ott}, J., \& {Tabatabaei}, F.
  2015, \apj, 800, 92

\bibitem[{{McClure-Griffiths} {et~al.}(2001){McClure-Griffiths}, {Green},
  {Dickey}, {Gaensler}, {Haynes}, \& {Wieringa}}]{McClure2001}
{McClure-Griffiths}, N.~M., {Green}, A.~J., {Dickey}, J.~M., {et~al.} 2001,
  \apj, 551, 394

\bibitem[{{Montier} {et~al.}(2015{\natexlab{a}}){Montier}, {Plaszczynski},
  {Levrier}, {Tristram}, {Alina}, {Ristorcelli}, \& {Bernard}}]{Montier2015a}
{Montier}, L., {Plaszczynski}, S., {Levrier}, F., {et~al.} 2015{\natexlab{a}},
  \aap, 574, A135

\bibitem[{{Montier} {et~al.}(2015{\natexlab{b}}){Montier}, {Plaszczynski},
  {Levrier}, {Tristram}, {Alina}, {Ristorcelli}, {Bernard}, \&
  {Guillet}}]{Montier2015b}
---. 2015{\natexlab{b}}, \aap, 574, A136

\bibitem[{{M{\"u}ller} {et~al.}(2017){M{\"u}ller}, {Beck}, \&
  {Krause}}]{Muller2017}
{M{\"u}ller}, P., {Beck}, R., \& {Krause}, M. 2017, \aap, 600, A63

\bibitem[{{Oppermann} {et~al.}(2015){Oppermann}, {Junklewitz}, {Greiner},
  {En{\ss}lin}, {Akahori}, {Carretti}, {Gaensler}, {Goobar}, {Harvey-Smith},
  {Johnston-Hollitt}, {Pratley}, {Schnitzeler}, {Stil}, \&
  {Vacca}}]{Oppermann2015}
{Oppermann}, N., {Junklewitz}, H., {Greiner}, M., {et~al.} 2015, \aap, 575,
  A118

\bibitem[{{O'Sullivan} {et~al.}(2012){O'Sullivan}, {Brown}, {Robishaw},
  {Schnitzeler}, {McClure-Griffiths}, {Feain}, {Taylor}, {Gaensler},
  {Landecker}, {Harvey-Smith}, \& {Carretti}}]{OSullivan2012}
{O'Sullivan}, S.~P., {Brown}, S., {Robishaw}, T., {et~al.} 2012, \mnras, 421,
  3300

\bibitem[{{O'Sullivan} {et~al.}(2013){O'Sullivan}, {Feain},
  {McClure-Griffiths}, {Ekers}, {Carretti}, {Robishaw}, {Mao}, {Gaensler},
  {Bland-Hawthorn}, \& {Stawarz}}]{OSullivan2013}
{O'Sullivan}, S.~P., {Feain}, I.~J., {McClure-Griffiths}, N.~M., {et~al.} 2013,
  \apj, 764, 162

\bibitem[{{Planck Collaboration} {et~al.}(2016){Planck Collaboration}, {Adam},
  {Ade}, {Alves}, {Ashdown}, {Aumont}, {Baccigalupi}, {Banday}, {Barreiro},
  {Bartolo}, {Battaner}, {Benabed}, {Benoit-L{\'e}vy}, {Bernard}, {Bersanelli},
  {Bielewicz}, {Bonavera}, {Bond}, {Borrill}, {Bouchet}, {Boulanger}, {Bucher},
  {Burigana}, {Butler}, {Calabrese}, {Cardoso}, {Catalano}, {Chiang},
  {Christensen}, {Colombo}, {Combet}, {Couchot}, {Crill}, {Curto}, {Cuttaia},
  {Danese}, {Davis}, {de Bernardis}, {de Rosa}, {de Zotti}, {Delabrouille},
  {Dickinson}, {Diego}, {Dolag}, {Dor{\'e}}, {Ducout}, {Dupac}, {Elsner},
  {En{\ss}lin}, {Eriksen}, {Ferri{\`e}re}, {Finelli}, {Forni}, {Frailis},
  {Fraisse}, {Franceschi}, {Galeotta}, {Ganga}, {Ghosh}, {Giard}, {Gjerl{\o}w},
  {Gonz{\'a}lez-Nuevo}, {G{\'o}rski}, {Gregorio}, {Gruppuso}, {Gudmundsson},
  {Hansen}, {Harrison}, {Hern{\'a}ndez-Monteagudo}, {Herranz}, {Hildebrandt},
  {Hobson}, {Hornstrup}, {Hurier}, {Jaffe}, {Jaffe}, {Jones}, {Juvela},
  {Keih{\"a}nen}, {Keskitalo}, {Kisner}, {Knoche}, {Kunz}, {Kurki-Suonio},
  {Lamarre}, {Lasenby}, {Lattanzi}, {Lawrence}, {Leahy}, {Leonardi}, {Levrier},
  {Liguori}, {Lilje}, {Linden-V{\o}rnle}, {L{\'o}pez-Caniego}, {Lubin},
  {Mac{\'{\i}}as-P{\'e}rez}, {Maggio}, {Maino}, {Mandolesi}, {Mangilli},
  {Maris}, {Martin}, {Mart{\'{\i}}nez-Gonz{\'a}lez}, {Masi}, {Matarrese},
  {Melchiorri}, {Mennella}, {Migliaccio}, {Miville-Desch{\^e}nes}, {Moneti},
  {Montier}, {Morgante}, {Munshi}, {Murphy}, {Naselsky}, {Nati}, {Natoli},
  {N{\o}rgaard-Nielsen}, {Oppermann}, {Orlando}, {Pagano}, {Pajot}, {Paladini},
  {Paoletti}, {Pasian}, {Perotto}, {Pettorino}, {Piacentini}, {Piat},
  {Pierpaoli}, {Plaszczynski}, {Pointecouteau}, {Polenta}, {Ponthieu}, {Pratt},
  {Prunet}, {Puget}, {Rachen}, {Reinecke}, {Remazeilles}, {Renault}, {Renzi},
  {Ristorcelli}, {Rocha}, {Rossetti}, {Roudier}, {Rubi{\~n}o-Mart{\'{\i}}n},
  {Rusholme}, {Sandri}, {Santos}, {Savelainen}, {Scott}, {Spencer},
  {Stolyarov}, {Stompor}, {Strong}, {Sudiwala}, {Sunyaev}, {Suur-Uski},
  {Sygnet}, {Tauber}, {Terenzi}, {Toffolatti}, {Tomasi}, {Tristram}, {Tucci},
  {Valenziano}, {Valiviita}, {Van Tent}, {Vielva}, {Villa}, {Wade}, {Wandelt},
  {Wehus}, {Yvon}, {Zacchei}, \& {Zonca}}]{Planck2016b}
{Planck Collaboration}, {Adam}, R., {Ade}, P.~A.~R., {et~al.} 2016, \aap, 596,
  A103

\bibitem[{{Robishaw} {et~al.}(2015){Robishaw}, {Green}, {Surcis}, {Vlemmings},
  {Richards}, {Etoka}, {Bourke}, {Fish}, {Gray}, {Imai}, {Kramer}, {McBride},
  {Momjian}, {Sarma}, \& {Zijlstra}}]{Robishaw2015}
{Robishaw}, T., {Green}, J., {Surcis}, G., {et~al.} 2015, Advancing
  Astrophysics with the Square Kilometre Array (AASKA14), 110

\bibitem[{{Robitaille} \& {Scaife}(2015)}]{Robitaille2015}
{Robitaille}, J.-F., \& {Scaife}, A.~M.~M. 2015, \mnras, 451, 372

\bibitem[{{Robitaille} {et~al.}(2017){Robitaille}, {Scaife}, {Carretti},
  {Gaensler}, {McEwen}, {Leistedt}, {Haverkorn}, {Bernardi}, {Kesteven},
  {Poppi}, \& {Staveley-Smith}}]{Robitaille2017}
{Robitaille}, J.-F., {Scaife}, A.~M.~M., {Carretti}, E., {et~al.} 2017, \mnras,
  468, 2957

\bibitem[{{Saikia} \& {Salter}(1988)}]{Saikia1988}
{Saikia}, D.~J., \& {Salter}, C.~J. 1988, \araa, 26, 93

\bibitem[{{Simmons} \& {Stewart}(1985)}]{Simmons1985}
{Simmons}, J.~F.~L., \& {Stewart}, B.~G. 1985, \aap, 142, 100

\bibitem[{{Soler} {et~al.}(2013){Soler}, {Hennebelle}, {Martin},
  {Miville-Desch{\^e}nes}, {Netterfield}, \& {Fissel}}]{Soler2013}
{Soler}, J.~D., {Hennebelle}, P., {Martin}, P.~G., {et~al.} 2013, \apj, 774,
  128

\bibitem[{{Stokes}(1852)}]{Stokes1852}
{Stokes}, G.~G. 1852, Trans. Cambridge Philos. Soc., 9, 399

\bibitem[{{Sun} {et~al.}(2014){Sun}, {Gaensler}, {Carretti}, {Purcell},
  {Staveley-Smith}, {Bernardi}, \& {Haverkorn}}]{Sun2014}
{Sun}, X.~H., {Gaensler}, B.~M., {Carretti}, E., {et~al.} 2014, \mnras, 437,
  2936

\bibitem[{{Sun} {et~al.}(2015){Sun}, {Landecker}, {Gaensler}, {Carretti},
  {Reich}, {Leahy}, {McClure-Griffiths}, {Crocker}, {Wolleben}, {Haverkorn},
  {Douglas}, \& {Gray}}]{Sun2015}
{Sun}, X.~H., {Landecker}, T.~L., {Gaensler}, B.~M., {et~al.} 2015, \apj, 811,
  40

\bibitem[{{Van Eck} {et~al.}(2017){Van Eck}, {Haverkorn}, {Alves}, {Beck}, {de
  Bruyn}, {En{\ss}lin}, {Farnes}, {Ferri{\`e}re}, {Heald}, {Horellou},
  {Horneffer}, {Iacobelli}, {Jeli{\'c}}, {Mart{\'{\i}}-Vidal}, {Mulcahy},
  {Reich}, {R{\"o}ttgering}, {Scaife}, {Schnitzeler}, {Sobey}, \&
  {Sridhar}}]{VanEck2017}
{Van Eck}, C.~L., {Haverkorn}, M., {Alves}, M.~I.~R., {et~al.} 2017, \aap, 597,
  A98

\bibitem[{{Vidal} {et~al.}(2016){Vidal}, {Leahy}, \& {Dickinson}}]{Vidal2016}
{Vidal}, M., {Leahy}, J.~P., \& {Dickinson}, C. 2016, \mnras, 461, 698

\bibitem[{{Wardle} \& {Kronberg}(1974)}]{Wardle1974}
{Wardle}, J.~F.~C., \& {Kronberg}, P.~P. 1974, \apj, 194, 249

\bibitem[{{Zaldarriaga} \& {Seljak}(1997)}]{Zaldarriaga1997}
{Zaldarriaga}, M., \& {Seljak}, U. 1997, \prd, 55, 1830

\bibitem[{{Zhang} {et~al.}(2016){Zhang}, {Lazarian}, {Lee}, \&
  {Cho}}]{Zhang2016}
{Zhang}, J.-F., {Lazarian}, A., {Lee}, H., \& {Cho}, J. 2016, \apj, 825, 154

\end{thebibliography}
\end{document}